\documentstyle{mn}
\input epsf
\def\plotone#1{\centering \leavevmode
\epsfxsize=\columnwidth \epsfbox{#1}}

\def\apj{ApJ}

\def\aap{A\&A}                
          
\def\aaps{A\&AS}

\def\mnras{MNRAS}

\def\pasj{PASJ}

\def\nat{Nature}

\def\deg{$^{\circ}$}

\title[Polarized emission from the Sgr B2 cloud]{Polarization of  X--ray emission from the Sgr B2 cloud}

\author[Churazov, Sunyaev and Sazonov]{E.~Churazov$^{1,2}$, R.~Sunyaev$^{1,2}$, S.~Sazonov$^{1,2}$\\
$^1$ Max-Planck-Institut f\"ur Astrophysik, Karl-Schwarzschild-Strasse 1, 85741
Garching, Germany\\
$^2$ Space Research Institute (IKI), Profsoyuznaya 84/32, Moscow 117810, 
Russia\\
}

\pagerange{\pageref{firstpage}--\pageref{lastpage}}
\pubyear{2001}

\begin{document}
\maketitle

\label{firstpage}
\begin{abstract}
The Sgr B2 giant molecular cloud is claimed to be an ''X--ray
reflection nebula'' -- the reprocessing site of a powerful flare of
the Sgr A* source, occurred few hundred years ago. The shape of the
X--ray spectrum and the strength of the iron fluorescent line support
this hypothesis. We argue that the most clean test of the origin of
X--rays from Sgr B2 would be a detection of polarized
emission from this source.
\end{abstract}

\begin{keywords}
Polarization -- scattering -- ISM: individual: Sgr B -- Galaxy: centre
-- X-rays: general
\end{keywords}

%

\sloppypar

\section{Introduction}
ASCA observations of the Sgr B2 giant molecular cloud revealed very
hard X--ray spectrum with a very prominent iron fluorescent line at
6.4 keV (Koyama et al., 1996). This discovery
provided an important confirmation of the hypothesis of Sunyaev,
Markevitch and Pavlinsky (1993) that the diffuse emission from
the giant molecular clouds in the Galactic Centre region is at least
partly due to reprocessed emission of a powerful X--ray flare 
from the supermassive black hole Sgr A*. The geometry of the problem
suggests that such flare 
could have happened few hundred years ago. The morphology and the
spectrum of the reprocessed emission have been modeled by Sunyaev \&
Churazov 1998 and Murakami et al. 2000. Recent SAX (Sidoli et al.,
2001) and Chandra data (Murakami,
Koyama, Maeda 2001a) are broadly consistent with the assumption that
reprocessed (reflected) emission is due to the past flare from Sgr
A*. Main observational arguments in favor of this interpretation are:
\begin{itemize}
\item Remarkably hard shape of the Sgr B2 X--ray spectrum
\item Extremely high flux in the neutral iron fluorescent line at 6.4
keV (equivalent width $\sim$1--2 keV).
\item The side of the cloud towards Sgr A* is brighter in X--rays than
the opposite side.
\end{itemize}
We argue below that the most convincing test of the origin of the
diffuse X--ray emission from the Sgr B2 cloud would be a detection of
polarized emission from this object. Recent progress in the
development of the X--ray polarimeters for space missions (Costa et
al., 2001) implies a drastic increase of the detector sensitivity and
makes the Sgr B2 cloud a natural target for the polarimetric studies.

\section{Simulations}
We consider a uniform cloud with the radius of 10 pc and the Thomson
optical depth of 0.5 exposed to the hard (power law photon index
$\alpha=1.8$) unpolarized X--ray radiation from an external source. As
we argue below the particular values of the cloud size and its optical
depth do not strongly affect the degree of the polarization of the
reprocessed/reflected emission. 
The external source was assumed to be steady on the time scales
comparable with the light crossing time of the cloud (the morphology
of the reprocessed emission from the cloud illuminated by a short flare
is considered in Sunyaev \& Churazov 1998). The cloud 
consists of hydrogen (in molecular form), helium and admixture of heavy
elements with the solar abundance. The reprocessed radiation was
calculated via Monte--Carlo method. The following three processes have
been taken into account: i) photoelectric absorption, ii) Compton
scattering by bound electrons and iii) fluorescent emission of heavy
elements. The cross sections for Compton scattering by electrons in
hydrogen molecules and helium atoms were taken from Sunyaev \&
Churazov 1996, Vainshtein, Sunyaev \& 
Churazov 1998 and Sunyaev, Uskov \& Churazov 1999. The fluorescent
yields and the energies of the fluorescent lines were taken from Kaastra
\& Mewe 1993. 

\begin{figure} 
\plotone{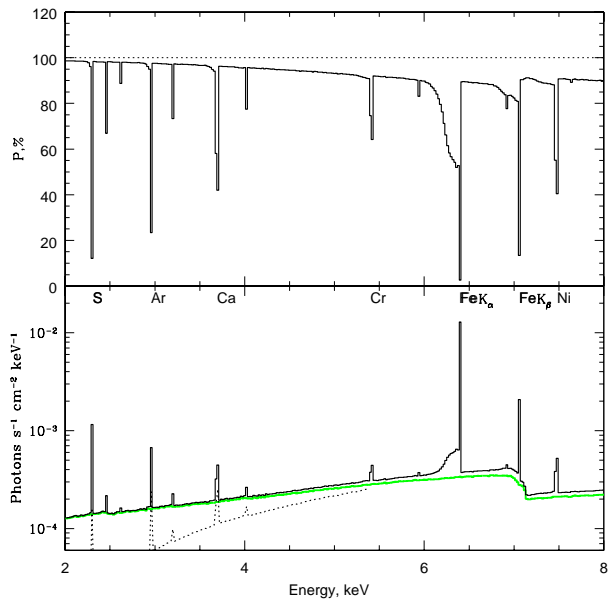}
\caption{{\bf Bottom panel:} A reflected energy spectrum (the solid
line) of a spherical cloud 
exposed to hard X--ray emission from an external source. The cloud is
assumed to be at the same distance from the 
observer as the primary source. The spectrum is shown with the energy
resolution of 20 eV. 
The brightest fluorescent lines of the most abundant elements
are marked. Note that for Ca, Cr and Ni the two components of the
$K_\alpha$ lines ($K_{\alpha_1}$ and $K_{\alpha_2}$) fell into adjacent
energy bins causing the lines to look broader. 
The dotted line shows the suppression of the low energy part of
the spectrum due to a photoabsorption over the line of sight for the
hydrogen column density of $5~10^{22}~cm^{-2}$.
The grey line shows the Stokes 
Q parameter, defined relative to the direction from Sgr A* to Sgr B2,
as a function of energy (the U parameter is equal to zero).  {\bf Top
panel:} Degree of 
polarization in the reflected spectrum. In the simulated geometry an
average scattering angle of the primary radiation is close to
90\deg and the continuum radiation is almost completely polarized. The
degree of continuum polarization slowly decreases with energy due to
the increasing contribution from multiple scatterings.
 The fluorescent lines are produced in the cloud itself and are emitted
isotropically. As a result degree of polarization drops strongly at the
energies of the lines. 
\label{sp}
}
\end{figure}

\begin{figure} 
\plotone{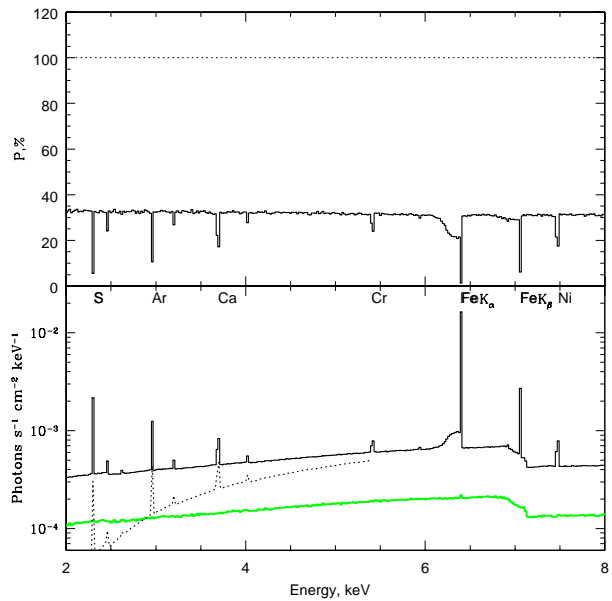}
\caption{The same as in Fig.1, but for the scattering cloud located
further 100 pc away from the observer than the illuminating
source. Projected distance of the cloud from the source is 100 pc --
the same as was assumed in Fig.1. An average scattering angle in this
case is 135\deg and the degree of polarization is close to 33\%. The
same degree of polarization is expected for the cloud located
by 100 pc closer to the observer than the illuminating source, when an
average scattering angle in $\sim$45\deg.
\label{sp135}
}
\end{figure}

In particular case of the Sgr B2 cloud (as a reflection nebula) and Sgr
A* (as an external source) the projected distance of $\sim$100 pc is
known. The cloud was assumed to be located at the same distance as Sgr
A* and outgoing photons, used for calculating degree of polarization,
were accumulated over small solid angle towards the observer's
direction. The resulting spectrum and the degree of polarization are
shown in Fig.\ref{sp}. For comparison the energy spectrum and the
degree of polarization have been calculated for the case when
scattering cloud is located at the same projected distance, but
further 100 pc away from the observer than the illuminating source
(Fig.\ref{sp135}).  

\section{Discussion and conclusions}
\subsection{Energy spectrum}
The spectra shown in Fig.\ref{sp} and \ref{sp135} have a typical
``reflection'' shape 
of the continuum and a set of strong fluorescent lines of the
astrophysically abundant elements like  S, Ar, Ca, Fe and Ni. At
low energies (below $\sim$7 keV) the shape of the continuum emission
is determined by the ratio of the scattering cross section to the
photoabsorption cross section. Due to the coherent scattering by the
electrons bound in hydrogen molecules and helium atoms the scattered
continuum is higher (by a factor of 2 at lowest energies) than the
reflected continuum calculated for free electrons. The ``Compton
shoulders'' on the red side of the strong lines are smooth and do not
show sharp features typical for scattering by the free cold electrons
(Sunyaev and Churazov 1996). For the iron 6.4 keV line the flux in the
shoulder is $\sim$10--20 \% of the line flux. 

Additional modification of the spectra shown in Fig.\ref{sp} and
\ref{sp135} should be due to interstellar absorption along the line of
sight. For the hydrogen column densities typical for the Galactic
Centre region (i.e. $N_H$ of $\sim 5~10^{22}~cm^{-2}$) 
the emission below 4-5 keV will be strongly suppressed.

\subsection{Unpolarized radiation of the primary source}
As is clear from Fig.\ref{sp} the continuum emission is strongly
polarized, while the emission in the fluorescent lines is not. This is
of course a natural result given the geometry of the problem. For the
neutral (or molecular) gas with solar abundance of heavy elements the
first scattering gives the dominant contribution to the
reflected continuum at energies below $\sim$7 keV.  For a Thomson
thick cloud the contribution from second scattering is 
attenuated by a factor $\delta(E)=\sigma_T/\sigma_{Ph}(E)\ll 1$, where
$\sigma_T$ is the Thomson cross section and $\sigma_{Ph}(E)$ is the
photoelectric cross section at given energy $E$. For a cloud with a
very small optical depth $\tau_T\ll 1$ the suppression factor for
second scattering is 
$\delta(E)=\tau_T$. Therefore contribution of second scattering is
almost always modest (at least at low energies) and reasonable 
estimate of the degree of polarization can be made under single
scattering approximation. The scattering cross section is proportional to
the factor $({\bf e_1}\cdot {\bf e_2})^2$, where ${\bf e_1}$ and
${\bf e_2}$ are the polarization vectors of the photon before and
after the scattering. Since the angular size of the reflecting cloud
(as seen from the illuminating source) is small then the degree of
polarization $P$ is related to the scattering angle $\theta$ via
well known expression: 
\begin{eqnarray} 
{\rm P=\frac{1- \mu^2}{1+\mu^2}},
\label{pol}
\end{eqnarray}
where $\mu=cos \theta$.
In particular case of the scattering by $\sim$90\deg simulated above
(Fig.\ref{sp}), the degree of polarization should be close to
unity. The deviation of the degree of polarization from unity 
in simulations is  caused mainly by the contribution of the second and
subsequent scatterings to the total spectrum. For the
second simulation (Fig.\ref{sp135}) the scattering angle is around
135\deg and the degree of polarization is close to 33\% in agreement
with the expression (\ref{pol}) for Rayleigh law of scattering.

Using again the argument that the angular size of the cloud (as seen
from the primary source) is relatively small one can treat the
incident radiation as a narrow 
unpolarized beam and get simple estimate (see Appendix) of the degree
of polarization for photons experienced two or more scatterings:
\begin{eqnarray} 
{\rm P_n=\frac{1-
\mu^2}{1+\mu^2+\frac{20}{15}(\left[\frac{10}{7}\right ]^{n-1}-1)}},
\label{poln}
\end{eqnarray}
where $n$ is the number of scatterings. From this equation it is clear
that the degree of polarization rather slowly decreases with the
increase of the number of scatterings. Therefore even although the
degree of polarization decreases with energy (see Fig.\ref{sp}) due to
increasing contribution from multiple scatterings, the reflected
continuum radiation remains nevertheless strongly polarized at all
energies. The direction of polarization remains the same
(perpendicular to the direction from the primary source to the cloud)
after any number of scatterings.

Fluorescent photons, which are born inside the cloud and are emitted
isotropically should not be polarized as indeed seen in the simulations. The 
Compton shoulder of the fluorescent lines can be polarized for some
specific geometries of the scattering clouds. E.g. for an optically
thin, strongly elongated cloud (in the direction perpendicular to the
line of sight) the shoulder can be strongly polarized. But in simple
cases like considered here the shoulder can be polarized
only weakly (compared to the continuum). 

Thus sensitive polarimetric observations of the Sgr B2 cloud should
reveal strongly polarized continuum and unpolarized
fluorescent lines. The polarization direction must be perpendicular to the
direction from the reflection nebula towards the primary source. Actual
degree of polarization is a function of the scattering angle and
therefore can be used to infer the mutual location of the Sgr B2 cloud
and the Sgr A* source along the line of sight. For Sgr B2 
cloud the total continuum flux in the 4-10 keV band is 1.5--2
$10^{-4}~ph~s^{-1} cm^{-2}$ as observed by ASCA. For a 4\% efficient
detector (see Costa et al., 2001) placed 
under $\sim$ 1000 $cm^2$ effective area mirror this source would yield about
$6~10^{-3}~cnt~s^{-1}$ in polarized flux. Therefore (assuming that the
detector background can be neglected) a significant detection can be
achieved in few $10^5$ s.  The Sgr B2 is not the only object which is
suspected to be the ``reflection nebula''. Murakami et al., 2001b
found strong fluorescent line emission from another rich molecular
complex -- Sgr C. As argued by Sunyaev, Markevitch, \& Pavlinsky (1993)
and Cramphorn \& Sunyaev (2001) many other reflection sites could
contribute to the diffuse emission in the Galactic Centre
region and the Galactic Ridge. X--ray polarimetric studies would then
allow one to reconstruct 
real 3-dimensional positions of the scattering clouds. Note however
that (in the first scattering approximation) the degree of
polarization is the same for the scattering angles of $\theta$ and
$\pi-\theta$, which correspond to the cloud shifted with respect to
the primary source towards or away from the observer by the same
distance. In order to break this degeneracy one have to use
detailed information on shape of the reflected continuum at low
energies. For smaller scattering angles (the scattering cloud is closer
to the observer than the primary source) the more distant side of the
cloud is exposed to primary radiation and photoabsorption should
suppress the reprocessed radiation at low energies. 

\subsection{Polarized radiation of the primary source}
More confusing can be the case when the emission of the primary source
is already polarized. For example this can be Sgr A* jet X--ray
emission or emission from hot optically thin quasi-flat accretion disk
(Sunyaev \& Titarchuk, 1985). The incident light is assumed to consist
of unpolarized component with the intensity $a0$ and the polarized
component with the 
intensity $a$. The degree of polarization of the initial radiation is
thus: $P_0=a/(a+a0)$. 
The degree of polarization $P$ of the scattered continuum radiation (in the
one scattering approximation) can then be easily calculated:
\begin{eqnarray} 
{\rm
P=\frac{\sqrt{(1-\mu^2+P_0(1+\mu^2)cos2\phi)^2+4P_0^2\mu^2sin^22\phi}}{1+\mu^2+P_0(1-\mu^2)cos
2\phi}},
\label{polj}
\end{eqnarray}
where $\phi$ is the angle between the direction of polarization of the
primary radiation and the perpendicular to the scattering plane.
This would make  the determination of
the cloud position with respect to the primary source more
complicated. However the continuum radiation will remain polarized in
majority of cases and therefore the hypothesis of the Sgr B2 as the
``reflection nebula'' 
can still be tested with the polarimetric observations. Polarization
of the incident radiation will also affect shape of the continuum spectrum
because it will modify the contributions of single and multiple
scattered photons to the total spectra. The equivalent width of the
fluorescent lines will also change. E.g. consider the single
scattering approximation for the 90\deg scattering. For the
completely polarized primary radiation the scattered continuum will be
attenuated by a factor $cos^2\phi$, while the intensity of the
fluorescent line will not be affected. Thus the equivalent width of
the line is:
\begin{eqnarray}
EW(\phi)=\frac{EW_0}{2cos^2\phi} ,
\end{eqnarray}
where $EW_0$ is the equivalent width of the line in the case of
unpolarized primary radiation. Therefore when the direction of the primary
radiation polarization is in the scattering 
plane the contribution from the first scattering to the continuum
spectrum is zero and the equivalent width of the iron line will be
much larger than for unpolarized primary radiation. In reality
presence of unpolarized component in the primary radiation and 
contribution from multiple scatterings will limit the maximal value of the
equivalent width. 

Finally we note that if the X--ray flare from the Sgr A* source was
short compared to the light crossing time of the Sgr B2 cloud (see
Sunyaev \& Churazov 1998) then the time delay for photons undergoing two
or more scatterings has to be taken into account when calculating
the energy spectrum and the degree of polarization. In the first scattering
approximation the results are unchanged: the continuum is
polarized in accordance with the scattering angle and the flux in the
fluorescent lines is completely unpolarized. 

\section{Acknowledgements}
We acknowledge helpful discussions with W. Forman. 
This research was partially supported by the Russian Foundation for
Basic Research (projects 00-02-16681 and 00-15-96649) and by the
program of the  Russian Academy of Sciences "Astronomy (Nonstationary 
astronomical objects)"

\appendix
\section{Polarization of a narrow beam after several scatterings.}
Consider the narrow beam of unpolarized light in the scattering medium
with the Rayligh scattering phase function. The light
after $n$ scattering can be represented as a two component vector:
\begin{eqnarray} 
{\rm \left ( 
\begin{array}{c}
I_{n,\parallel}(\mu)\\
I_{n,\bot}(\mu)
\end{array}
\right )},
\end{eqnarray}
where $\mu$ is the cosine of the angle with respect to the beam axis and $I_{n,\bot}(\mu)$ and $I_{n,\parallel}(\mu)$ are the intensities of light
with the directions of polarization perpendicular and parallel to the
plane formed by the initial beam and the direction of the photon. From
the symmetry of the problem it is clear that only two components are
needed to fully describe the polarization. The change of the polarization after
additional one scattering is then governed by the simple
transformation (e.g. Chandrasekhar, 1950):
\begin{eqnarray}
\label{ach}
\left (
\begin{array}{c}
I_{n+1,\parallel}(\mu) \\
I_{n+1,\bot}(\mu)
\end{array}
\right ) =\frac{3}{8}\times ~~~~~~~~~~~~~~~~~~~~~~~\\
\int_{-1}^{+1}
\left ( 
\begin{array}{cc}
2(1-\mu^2)(1-\mu'^2)+\mu^2\mu'^2 & \mu^2 \\
\mu'^2 & 1 
\end{array}
    \right ) \left (
\begin{array}{c}
I_{n,\parallel}(\mu')\\
I_{n,\bot}(\mu')
\end{array}
\right ) 
d\mu' \nonumber
\end{eqnarray}
The initial values of $I$ for a narrow beam of unpolarized light are
obviousely: 
\begin{eqnarray} 
{\rm \left ( 
\begin{array}{c}
I_{0,\parallel}(\mu)\\
I_{0,\bot}(\mu)
\end{array}
\right )=
\frac{1}{2}\left ( 
\begin{array}{c}
\delta(\mu-1)\\
\delta(\mu-1)
\end{array}
\right )
},
\label{am}
\end{eqnarray}
where $\delta$ is a Dirac function. From (\ref{ach}) it is clear that
after any number of scatterings the intensities $I$ can be written in
the form: 
\begin{eqnarray} 
\left (
\begin{array}{c}
I_{n,\parallel}(\mu)\\
I_{n,\bot}(\mu)
\end{array}
\right ) = \left (
\begin{array}{c}
a_n+b_n \mu^2\\
c_n
\end{array}
\right ).
\label{a2}
\end{eqnarray}
E.g. for the radiation scattered once we obtain, by substituting
(\ref{am}) into (\ref{ach}): $I_{1,\parallel}=3/8\mu^2$,
$I_{1,\bot}=3/8$. I.e. $a_1=0$; $b_1=3/8$; $c_1=3/8$. 

Inserting the expression (\ref{a2}) into (\ref{ach}) one gets the recursive
relation for $a_n$, $b_n$ and $c_n$:
\begin{eqnarray} 
\left (
\begin{array}{c}
a_{n+1}\\
b_{n+1}\\
c_{n+1}
\end{array}
\right ) = \left (
\begin{array}{c}
a_n+\frac{1}{5}b_n\\
-\frac{3}{4}a_n-\frac{1}{20}b_n+\frac{3}{4}c_n \\
\frac{1}{4}a_n+\frac{3}{20}b_n+\frac{3}{4}c_n
\end{array}
\right ).
\label{arec}
\end{eqnarray}
From  (\ref{arec}) it follows that $c_{n+1}=a_{n+1}+b_{n+1}$. The
remaining two relations are then simplify to:
\begin{eqnarray} 
\left (
\begin{array}{c}
a_{n+1}\\
b_{n+1}
\end{array}
\right ) = \left (
\begin{array}{c}
a_n+\frac{1}{5}b_n\\
\frac{7}{10}b_n
\end{array}
\right ).
\label{arec2}
\end{eqnarray}
Using (\ref{arec2}) and known values of $a_1$ and $b_1$ we obtain the
explicit expression:
\begin{eqnarray} 
\left (
\begin{array}{c}
a_{n}\\
b_{n} \\
c_{n}
\end{array}
\right ) = \left (
\begin{array}{c}
\frac{1}{4} - \frac{1}{4} \left
(\frac{7}{10}\right )^{n-1} \\
\frac{3}{8}\left ( \frac{7}{10} \right )^{n-1}\\
\frac{1}{4}+\frac{1}{8}\left (\frac{7}{10}\right )^{n-1} 
\end{array}
\right ).
\label{arec3}
\end{eqnarray}
Finally by substituting (\ref{arec3}) into (\ref{a2})
we can calculate the degree of
polarization of photons experienced $n$ scatterings:
\begin{eqnarray} 
{\rm P_n=\frac{I_{n,\bot}-I_{n,\parallel}}{I_{n,\bot}+I_{n,\parallel}}=\frac{1-
\mu^2}{1+\mu^2+\frac{20}{15}(\left[\frac{10}{7}\right ]^{n-1}-1)}},
\label{afin}
\end{eqnarray}

\label{lastpage}

\end{document}